\documentclass[aps,preprintnumbers,twocolumn,superscriptaddress,nofootinbib]{revtex4}

\usepackage{bm,color,xcolor}
\usepackage{slashed}
\usepackage{graphicx}
\usepackage{epsfig,latexsym,cancel,amssymb,adjustbox}
\usepackage{epstopdf}
\usepackage{hyperref}
\usepackage{hyphenat}
\usepackage{multirow}
\usepackage[normalem]{ulem}

\newcommand{\GeV}{\ensuremath{\mathrm{GeV}}}
\newcommand{\TeV}{\ensuremath{\mathrm{TeV}}}

\newcommand{\beq}{\begin{equation}}
\newcommand{\eeq}{\end{equation}}
\newcommand{\bea}{\begin{eqnarray}}
\newcommand{\eea}{\end{eqnarray}}

\usepackage{amsmath}
\usepackage{physics}
\usepackage{array}
\usepackage{tikz}
\usepackage{yfonts}
\usepackage{eufrak}
\usepackage{siunitx}
\usepackage{ragged2e}
\usepackage{caption}
\usetikzlibrary{arrows,positioning,snakes}

\begin{document}

\begin{flushright}
\end{flushright}

\title{\bf \Large LHC Shines on Positivity}

\preprint{UMN-TH-4514/25}

\author{Zhen Liu}
\thanks{\scriptsize \!\! \href{https://orcid.org/0000-0002-3143-1976}{0000-0002-3143-1976}}
\affiliation{\it School of Physics and Astronomy, University of Minnesota, Minneapolis, MN 55455, USA}
\author{Kun-Feng Lyu}
\thanks{\scriptsize \!\! \href{https://orcid.org/0000-0002-3291-1701}{0000-0002-3291-1701}}
\affiliation{\it Homer L. Dodge Department of Physics and Astronomy, University of Oklahoma, Norman, OK
73019, USA}

\author{Tong Arthur Wu}
\thanks{\scriptsize \!\! \href{https://orcid.org/0009-0002-9485-3938}{0009-0002-9485-3938}}
\affiliation{PITT PACC, Department of Physics and Astronomy,
University of Pittsburgh, 3941 O’Hara St., Pittsburgh, PA 15260, USA}

\begin{abstract}
We show that hadron colliders have an excellent reach for positivity tests on a class of diphoton operators. Due to the helicity selection rules, the relevant dimension-6 operators either do not contribute or are highly constrained by other experimental observables.
We show, for the first time, that the LHC can probe the positivity of the dimension-8 operators involving colored particles. The kinematic differential distributions of the diphoton final states are exploited to perform the $\chi^2$ analysis. Through a global fit, the effective scale for these operators can be inclusively probed up to around 2 TeV at HL-LHC and over 5 TeV at future 100 TeV FCC-hh at 95\% C.L., providing a powerful test of the positivity bounds up to multi-TeV scale.
\end{abstract}

\maketitle

\section{Introduction}

Effective Field Theory (EFT) provides a model-independent bridge between infrared (IR) and ultraviolet (UV) physics. Heavy states generate higher-dimensional operators suppressed by powers of a cutoff scale $\Lambda$. Under general assumptions of unitarity, causality, locality, scattering amplitudes satisfy positivity bounds that restrict combinations of Wilson coefficients; these bounds become nontrivial starting at dimension eight \cite{Adams:2006sv, Distler:2006if, Ian:2010, deRham:2017avq, deRham:2017zjm, Bellazzini:2017bkb, deRham:2018qqo,Zhang:2018shp,Arkani-Hamed:2020blm, Bellazzini:2020cot,Tolley:2020gtv,Fuks:2020ujk,Yamashita:2020gtt,Caron-Huot:2020cmc,Zhang:2020jyn,Li:2021lpe,deRham:2022hpx,Ghosh:2022qqq,Chen:2023bhu,Li:2022rag,Ye:2024rzr,Hong:2024fbl,Desai:2025alt}.
Measuring the corresponding Wilson coefficients and testing their compatibility with these positivity bounds thus offers a unique way to directly probe the fundamental principles of quantum field theory. Any signature of positivity bound violation would not only indicate new physics beyond the Standard Model (BSM), but also undermine the validity of those principles.

Testing positivity at colliders is typically challenging because lower-dimensional operators (notably dimension six) generically yield larger effects in the same observables. Nevertheless, helicity selection rules can render certain processes insensitive to dimension-6 contributions~\cite{Azatov:2016sqh}. The diphoton process is one such example, where the leading BSM effects arise from dimension-8 operators, and therefore provides a clean test of the positivity bounds. Building on proposals to test positivity at lepton colliders~\cite{Gu:2020ldn} and related extensions to hadron colliders~\cite{Gu:2023emi}, we study, for the first time, how the hadron colliders enable us to probe the positivity bounds in the quark and gluon sectors through the diphoton channel.

In this study, we identify the relevant dimension‑8 contact operators coupling two photons to quark pairs or gluons, derive helicity amplitudes, and apply positivity bounds to obtain sign constraints on linear combinations of Wilson coefficients. Phenomenologically, the amplitude induced by the quark-photon operators interferes with the SM tree amplitude, while the BSM amplitude generated from the gluon-photon operators interferes with the SM one‑loop box diagram. 
Although loop-suppressed, the gluon channel benefits from large gluon parton distribution functions (PDFs), which can enhance sensitivity.
We estimate collider sensitivity using kinematic differential distributions and a $\chi^2$ analysis focused on the interference signal (linear in the dimension‑8 coefficients). 
Our projections indicate sensitivity to effective scales of order $\sim2$ TeV at the HL‑LHC and above $\sim5$ TeV at a 100 TeV proton collider (95\% C.L.), up to which the validity of the positivity bounds can be tested. 

%%%%%%%%%%%%%%%%%%%%%%%%%%%%%%%%%%%%%%%%%
\section{Di-photon Processes}

The diphoton final state at hadron colliders is experimentally clean and theoretically well controlled \cite{ATLAS:2012yve,CMS:2012qbp,D0:2008hxb,CMS:2011bsw,CMS:2014mvm,MARINI20161973,ATLAS:2017ayi,CMS:2018dqv,Hamberg:1990np,Binoth:1999qq,KUMAR200945,Campbell:2011bn,Catani:2011qz,Cieri:2013pza,Mangano:2016jyj,Campbell:2016yrh,Kamenik:2016tve,Catani:2018krb,Catani2018DiphotonPA,Chawdhry:2021hkp,Cieri:2021fdb,Becchetti:2023yat}. At leading order, the dominant mechanism is quark-antiquark annihilation $q\bar q\to\gamma\gamma$. 
Gluon-initiated production in SM first arises at one loop via the quark box diagram, which is subleading compared to $q\bar q\to\gamma\gamma$. However, when interfering with new physics at tree level, this channel can also be sizable. 
Additional channels such as $gq\to\gamma\gamma j$ produce extra jets, but photon isolation and jet-veto cuts can suppress these contributions so that the two-photon final state remains the principal observable.

The higher-dimensional operators modify these processes either by correcting SM vertices or by introducing new contact interactions coupling two photons to quarks or gluons. 
Gauge invariance forbids the local photon–gluon contact operator at dimension six. 
The leading local contact interactions first arise at dimension eight. 
In the following, we first show that the leading dimension-6 effects do not interfere with the SM amplitude, and then comment on the higher-order contributions
before presenting the dimension‑8 operators and their phenomenology.

\subsection{Dimension-6 Operators}

At dimension-6 level the leading operators that modify quark–photon couplings are dipole-type operators, such as $(\bar q\sigma^{\mu\nu}u)\tilde{\varphi} B_{\mu\nu}$ and $(\bar q\sigma^{\mu\nu}u)\tau^I\tilde{\varphi} W^I_{\mu\nu}$ \cite{Grzadkowski:2010es}. A single insertion of such a dipole operator flips the fermion chirality relative to the SM tree-level amplitude, so it has light-quark-mass-suppressed interference with the dominant SM $q\bar q\to\gamma\gamma$ configurations at leading order.~\footnote{For all that matters at the LHC, all light quarks are massless, so use helicity and chirality equivalently for the rest of the paper.}
The analogous quark–gluon dipole operator $(\bar q\sigma^{\mu\nu}T^A u)\tilde{\varphi} G^A_{\mu\nu}$ 
contributes at NLO and suffers the same helicity suppression. 
The square of the amplitude proportional to dimension-6 Wilson coefficients can contribute to the total cross section, but is suppressed due to the following reasons. 
First, the amplitude square of the dipole operators scales as $|M_{\rm d6}|^2\sim v^2 s/\Lambda^4$, 
which has less energy growth compared to the interference term of dimension-8 operators $\mathrm{Re}(\mathcal{M}_{\rm SM}^*\mathcal{M}_{\rm d8})\sim s^2/\Lambda^4$. 
Second, those operators are strongly constrained by other measurements. 
The scales of the quark electric dipole operators can already be constrained to $\Lambda>2\,\mathrm{TeV}$ using electroweak precision measurements and $36\,\mathrm{fb}^{-1}$ of LHC data, while HL-LHC projection can extend these bounds to $4-5\,\mathrm{TeV}$~\cite{daSilvaAlmeida:2019cbr,Gauld:2024glt}. 
For the color dipole operators, Ref.~\cite{Hayreter:2013kba} showed that the scales are bounded by $2\,\mathrm{TeV}$ from $pp\to hX$ process.

Other dimension‑six operators (for example the CP‑even triple‑gauge operator $\epsilon^{IJK}W^I W^J W^K$ and Higgs–gauge operators) can in principle modify diphoton production through loop insertion, but many of these operators are tightly constrained by measurements of triple gauge couplings and Higgs precision observables \cite{Cirigliano:2019vfc,Panico:2017frx,Azatov:2017kzw,Biekotter:2021int,An:2018dwb,Bian:2015zha,Ellis:2020unq}.

For the gluon‑initiated channel $gg\to\gamma\gamma$, insertions of dipole operators into the one‑loop box also induce helicity flips and do not produce a leading interference with the SM box. Purely gluonic dimension‑six operators (e.g. $f^{ABC}G^AG^BG^C$) are typically suppressed in their contribution to this observable. 
We summarize the effective scale reach of the relevant dimension-6 operators in the appendix.
Therefore, the diphoton process provides a clean measurement on the dimension-8 operators and test on the positivity bounds, without contamination of dimension-6 contributions.

\subsection{Basic set of Dimension-8 Operators}
Having shown that dimension‑6 operators either do not interfere or are tightly constrained, we focus on the dimension‑8 operators relevant for diphoton production \cite{Li:2020gnx,Murphy:2020rsh}. These operators fall into two classes: (i) quark‑portal operators coupling a quark current to two electroweak field strengths, and (ii) four field strength operators coupling gluons to electroweak field strengths. Positivity bounds probe amplitudes that scale as $s^2$, hence operators with two or more derivatives enter.

The relevant quark‑portal operators are given by,
\begin{align}
    \mathcal{O}_{uW} &= \frac{1}{4} ( i \Bar{u}_p \gamma^{\{\mu} \overleftrightarrow{D} ^{\nu\}} u_r) W^I_{\mu \rho} W^{I \rho}_{\nu}, \\
    \mathcal{O}_{uB} &= \frac{1}{4}  ( i \Bar{u}_p \gamma^{\{\mu} \overleftrightarrow{D} ^{\nu\}} u_r) B_{\mu \rho} B^{\hspace{0.5em} \rho}_{\nu}, \\
    \mathcal{O}_{dW} &= \frac{1}{4}  ( i \Bar{d}_p \gamma^{\{\mu} \overleftrightarrow{D} ^{\nu\}} d_r ) W^I_{\mu \rho} W^{I \rho}_{\nu}, \\
    \mathcal{O}_{dB} &= \frac{1}{4}  ( i \Bar{d}_p \gamma^{\{\mu} \overleftrightarrow{D} ^{\nu\}} d_r ) B_{\mu \rho} B^{\hspace{0.5em} \rho}_{\nu}, \\
    \mathcal{O}_{qW} &= \frac{1}{4}  ( i \Bar{q}_p \gamma^{\{\mu} \overleftrightarrow{D} ^{\nu\}} q_r ) W^I_{\mu \rho} W^{I \rho}_{\nu}, \\
    \mathcal{O}_{qB} &= \frac{1}{4}  ( i \Bar{q}_p \gamma^{\{\mu} \overleftrightarrow{D} ^{\nu\}} q_r ) B_{\mu \rho} B^{\hspace{0.5em} \rho}_{\nu}, \\
    \mathcal{O}_{qBW} &= \frac{1}{4}  ( i \Bar{q}_p \gamma^{\{\mu} \tau^I \overleftrightarrow{D} ^{\nu\}} q_r ) B_{\mu \rho} W^{I \rho}_{\nu}  \ , 
\end{align}
where $I$ is the SU(2) adjoint index, $p$ and $r$ refer to the flavor indices. In the following, we work in the flavor-diagonal limit, and focus on the first generation. 
All the relevant operators above induce the same Lorentz structure to the process $q \Bar{q} \rightarrow \gamma\gamma$. 
The other set of CP-even operators describes the interactions between gluons and photons, from the combinations of four field strength tensors, given by
\begin{align}
    \mathcal{O}^{(1)}_{G^2 W^2} &=   (W^I_{\mu\nu}W^{I\mu\nu})(G^A_{\rho\sigma}G^{A\rho\sigma}) \ ,\\
    \mathcal{O}^{(2)}_{G^2 W^2} &=  (W^I_{\mu\nu}\widetilde{W}^{I\mu\nu})(G^A_{\rho\sigma}\widetilde{G}^{A\rho\sigma}) \ ,\\
    \mathcal{O}^{(3)}_{G^2 W^2} &=  (W^I_{\mu\nu}G^{A\mu\nu})(W^I_{\rho\sigma}G^{A\rho\sigma}) \ , \\
    \mathcal{O}^{(4)}_{G^2 W^2} &=  (W^I_{\mu\nu}\widetilde{G}^{A\mu\nu})(W^I_{\rho\sigma}\widetilde{G}^{A\rho\sigma})\ , \\
    \mathcal{O}^{(1)}_{G^2 B^2} &=  (B_{\mu\nu}B^{\mu\nu})(G^A_{\rho\sigma}G^{A\rho\sigma}) \ , \\
    \mathcal{O}^{(2)}_{G^2 B^2} &=  (B_{\mu\nu}\widetilde{B}^{\mu\nu})(G^A_{\rho\sigma}\widetilde{G}^{A\rho\sigma}) \ , \\
    \mathcal{O}^{(3)}_{G^2 B^2} &=  (B_{\mu\nu}G^{A\mu\nu})(B_{\rho\sigma}G^{A\rho\sigma})  \ , \\
    \mathcal{O}^{(4)}_{G^2 B^2} &=  (B_{\mu\nu}\widetilde{G}^{A\mu\nu})(B_{\rho\sigma}\widetilde{G}^{A\rho\sigma}) \ ,
\end{align}
where $I$ and $A$ are the SU(2) and SU(3)$_C$ adjoint indexes respectively. For the $\mathcal{O}^{(1)}$ and $\mathcal{O}^{(2)}$ operators, the gluon field and the electroweak field are separately contracted. While for the $\mathcal{O}^{(3)}$ and $\mathcal{O}^{(4)}$ operators, one gluon field tensor would contract with the electroweak field tensor. The two different operator combinations lead to distinct kinematics. 
In the following, we come to analyze the positivity bound of these operators and their observational effects at the LHC. 

\subsection{$q\bar q\rightarrow \gamma\gamma$ processes}
We first consider the quark‑initiated channel and turn on the seven quark‑portal operators with Wilson coefficients $c_i/\Lambda^4$. 
In the SM, the nonvanishing helicity amplitudes for massless quarks are $\mathcal{M}(q_L \Bar{q}_L \gamma^+ \gamma^-)$ and $\mathcal{M}(q_R \Bar{q}_R  \gamma^+ \gamma^-)$. We adopt the convention that all external momenta are outgoing and $+/-$ label the right/left-handed helicity modes. Considering the amplitude $q(p_1)+\bar{q}(p_2) \rightarrow \gamma(p_3)+\gamma(p_4)$, the contribution from the dimension‑8 quark operators to the amplitudes takes the compact spinor form
\begin{align}
    \mathcal{M} (q_L \Bar{q}_L \gamma^+ \gamma^-)_{\mathrm{d8}} 
    &= \frac{C_{q_L}}{\Lambda^4}[13][23]\langle 24 \rangle ^2 ,\\
    \mathcal{M} (q_R \Bar{q}_R  \gamma^+ \gamma^-)_{\mathrm{d8}} &=\frac{C_{q_R}}{\Lambda^4}\langle 14 \rangle\langle 24 \rangle[23] ^2 ,
\end{align}
where $q=u,d$, and the weak‑mixing combinations entering the photon coupling are
\begin{align}
    \hspace{-0.2em} C_{u_L} \! \!  = \! c_{qW}^{(8)}  \sin^2 \! \theta_W \! + \! c_{qB}^{(8)}  \cos^2 & \theta_W \! - \! c_{qBW}^{(8)} \! \cos \! \theta_W \! \sin \! \theta_W, \\
    \hspace{-0.2em} C_{d_L} \! \! = \! c_{qW}^{(8)}  \sin^2 \! \theta_W \! + \! c_{qB}^{(8)}  \cos^2 & \theta_W \! + \! c_{qBW}^{(8)} \! \cos \! \theta_W \! \sin \! \theta_W, \\
    C_{u_R} = c_{uW}^{(8)} \sin^2 \theta_W &+ c_{uB}^{(8)} \cos^2 \theta_W, \\
    C_{d_R} = c_{dW}^{(8)} \sin^2 \theta_W &+ c_{dB}^{(8)} \cos^2 \theta_W.
\end{align}
Summing over initial‑state helicities yields a partonic differential distribution proportional to the averaged coefficient $C_q=(C_{q_L}+C_{q_R})/2$.

Applying the forward‑limit positivity bound \cite{deRham:2017avq,deRham:2017zjm} to the crossed elastic process $q\gamma\to q\gamma$ yields the sign constraints $C_{q_L}\ge0$ and $C_{q_R}\ge0$ (and hence $C_q\ge0$).

\subsection{$g g \rightarrow \gamma\gamma$ process}
The CP-even gluonic operators yield the following helicity amplitudes for $g(p_1)+g(p_2) \rightarrow \gamma(p_3)+\gamma(p_4)$
\begin{align}
    \mathcal{M}(g^+ g^+ \gamma^+ \gamma^+)_\text{d8} 
    &= \frac{1}{\Lambda^4}  \Big(4 [12]^2[34]^2 (C_{g,1}-C_{g,2}) \nonumber \\
    & \hspace{-3em} +2([13]^2[24]^2 + [14]^2[23]^2)(C_{g,3}-C_{g,4})\Big)\delta^{a_1a_2}, \notag
     \\ 
    \mathcal{M}(g^+ g^+ \gamma^- \gamma^-)_\text{d8} 
    &= \frac{1}{\Lambda^4}4 [12]^2\langle34\rangle^2(C_{g,1}+C_{g,2})\delta^{a_1a_2} , \notag\\
    \mathcal{M}(g^+ g^- \gamma^+ \gamma^-)_\text{d8} 
    &= \frac{1}{\Lambda^4}2 [13]^2\langle24\rangle^2 (C_{g,3}+C_{g,4})\delta^{a_1a_2}, \notag  \\
    \mathcal{M}(g^+ g^- \gamma^- \gamma^+)_\text{d8} 
    &= \frac{1}{\Lambda^4}2 [14]^2\langle23\rangle^2 (C_{g,3}+C_{g,4})\delta^{a_1a_2},\notag
\end{align}
where $a_{1,2}$ are the SU(3) adjoint indices for gluons, and the electroweak–gluon combinations are defined as $C_{g,i}=\sin^2\theta_W\,c^{(i)}_{G^2W^2}+\cos^2\theta_W\,c^{(i)}_{G^2B^2}$. 
The contraction of the field-strength tensors enforces equal helicities for the two gauge fields.
Consequently, the Lorentz structure singles out the operators in different channels. $C_{g,1}$ and $C_{g,2}$ enter only in the $s$-channel and $C_{g_3}$ and $C_{g,4}$ only show up in $t$ and $u$ channel.

Imposing the positivity bound on the crossed process $g\gamma \rightarrow g\gamma$, we find that 
\begin{equation}
    C_{g,3} > 0, \quad C_{g,4}>0\, .
\end{equation}
There is no constraint on the other two Wilson coefficients $C_{g,1}$ and $C_{g,2}$, for the amplitudes from the two contact terms vanish at the forward limit $t = 0$.

\section{Probing the positivity bound at LHC}

In this section we evaluate the LHC sensitivity to the dimension‑8 operators \cite{Hays:2018zze,Guo:2019agy,Alioli:2020kez,Ellis:2020ljj,Guo:2020lim,Yang:2021pcf,Ellis:2018cos,Ellis:2021dfa} relevant for positivity tests. The full amplitude consists of the SM contribution plus the EFT contact term. Our baseline is performed at leading order in QCD and focused on the interference term linear in the dimension‑8 Wilson coefficients. 
Expanding the squared amplitude leads to
\begin{equation}
    |\mathcal{M}_\text{SM} + \mathcal{M}_\text{d8}|^2 =
   |\mathcal{M}_\text{SM}|^2 + 2 \text{Re} (\mathcal{M}^*_\text{SM}  \mathcal{M}_\text{d8})  \ .
\end{equation}
%
%In the baseline projections
Here we neglect the $|\mathcal{M}_\text{d8}|^2$ ($\Lambda^8$-suppressed) term and treat the interference Re(\(\mathcal{M}_\text{SM}^*\mathcal{M}_\text{d8}\)) as the signal, with $|\mathcal{M}_\text{SM}|^2$ the dominant background.
\footnote{Compared to the previous study on the $|\mathcal{M}_\text{d8}|^2$ signal for the gluonic operators \cite{Ellis:2018cos,Ellis:2021dfa}, not concerning positivity, we find compatible sensitivity.}

We focus on two primary channels, the quark-initiated process $q \bar{q} \rightarrow \gamma \gamma$ and the gluon-initiated process $g g \rightarrow \gamma \gamma$. 
The signal rate of $g g \rightarrow \gamma\gamma$, originating from the interference between the SM loop diagram and the tree-level diagram from dimension-8 operators, is generally much larger in relative size than that of the quark-initiated interference.
This originates from the fact that the interference term carries only a single loop factor, whereas the SM gluonic channel is doubly loop-suppressed.
While the majority of SM events are concentrated in the low $\hat{s}$ region, the contribution from dimension-8 operators—characterized by high-derivative interactions—is enhanced at high $\hat{s}$.

For the numerical computation, we exploit the {\tt FeynCalc} package~\cite{Shtabovenko:2023idz,Shtabovenko:2020gxv,Shtabovenko:2016sxi,Mertig:1990an} with interfaced {\tt LoopTools}~\cite{Hahn:1998yk}. The convoluted PDF is chosen to be {\tt nCTEQ15FullNuc\_1\_1} \cite{Kovarik:2015cma}. Both the renormalization scale and factorization scale are set to be $\sqrt{\hat{s}}$. We vary this scale from $\sqrt{\hat{s}}/2$ to $2\sqrt{\hat{s}}$ as an estimate of the theoretical uncertainty. 
In our analysis, 
we work at the leading order (tree level for $q\bar{q}\rightarrow\gamma\gamma$ and one-loop box diagram for the SM contribution to $gg\rightarrow\gamma\gamma$). While higher-order QCD corrections can be sizable, characterized by an O(1) K-factor \cite{Catani:2011qz,Catani2018DiphotonPA}, our primary goal is to highlight the impact of the new physics. Hence, we leave the precision QCD improvement for future work.

\begin{figure}[!t]
    \centering
        \includegraphics[width=0.7\linewidth]{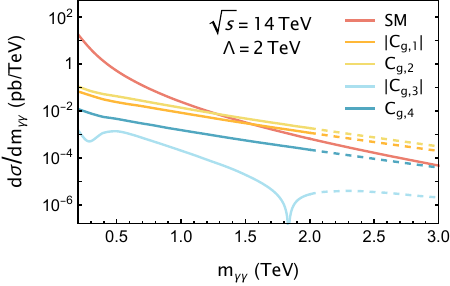}
    \\[3mm]
       \hspace{-1.1em} \includegraphics[width=0.7\linewidth]{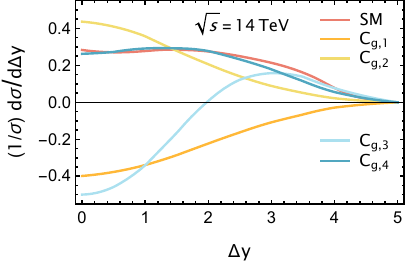}
    \caption{Top: invariant mass distribution of the SM background and the interference signal from gluon‑photon operators. Bottom: normalized $\Delta y$ distribution. Curves labelled $O_{g,i}$ show the interference term with the corresponding $C_{g,i}=1$ and $\Lambda=2\,$TeV. 
    }
    \label{fig:gg}
\end{figure}

As shown in the previous sections, there are six independent coefficients in total, namely $C_u, C_d$, $C_{g,1},C_{g,2},C_{g,3}$ and $C_{g,4}$. 
In the following, we first turn on each individual operator relevant to the positivity bound and display the constraint on the corresponding coefficient. Then all relevant operators are turned on simultaneously and the chi-square fitting is performed to derive the constraints on each Wilson coefficient. 
\begin{figure}[!t]
    \centering
        \includegraphics[width=0.7\linewidth]{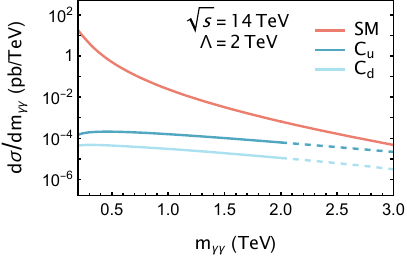}
        %\\[3mm]
        \includegraphics[width=0.65\linewidth]{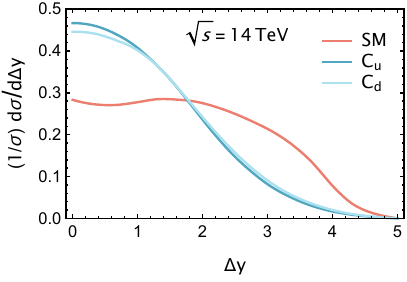}
    \caption{{Top}: invariant mass distribution of the SM background and the interference signal from quark‑photon operators. {Bottom}: normalized $\Delta y$
    %\ZL{y--rapidity of the dijet system}
    distribution. Curves labelled $O_u$ and $O_d$ show the interference term with only the up‑ or down‑quark operator switched on with $C_q=1$ and $\Lambda=2\,\TeV$.}
    \label{fig:qq}
\end{figure}
To fully exploit the available kinematic information, we first compute the 2D differential distribution $\dd\sigma/(\dd \sqrt{\hat{s}} \, \dd\Delta y)$ where $\Delta y$ is the rapidity difference of the two outgoing photons, to capture and characterize the angular dependence of different operators and processes. 
For illustration, the distributions $\dd\sigma/\dd \sqrt{\hat{s}}$ and $\dd\sigma/\dd \Delta y$ for SM and the operators are shown in Fig.~\ref{fig:qq} and~\ref{fig:gg}. The invariant mass $\sqrt{\hat{s}}$ is restricted to range from $200~ \GeV$ to 2 TeV and $\Delta y$ is limited from 0 to 5. 
The curves corresponding to $C_q$ and $C_{g,i}$ represent the interference terms, which are linear in  $C_q/\Lambda^4$ and $C_{g,i}/\Lambda^4$ respectively. For normalization, we adopt the benchmark values $C_q=C_{g,i}=1$ and $\Lambda = 2$ TeV. 
The following kinematic cuts are applied in the diphoton distributions,
\begin{equation}
    \eta(\gamma) < 2.5, \quad \quad p_T(\gamma) > 30 \GeV \ .
\end{equation}
As shown in the figures, the differential distribution in the SM case decreases with increasing diphoton invariant mass. 
In contrast, the distributions for $C_u$ and $C_d$ appear relatively flat. 
This behavior arises from the compensating effects of the dimension-8 operators, which enhance the high $\hat{s}$ region. 
The contribution from $C_u$  is several times larger than that from $C_d$, reflecting the larger PDF for 
$u$-quarks compared to 
$d$-quarks.
For the gluonic operators, the curves for $C_{g,1}$ and $C_{g,2}$ in Fig.~\ref{fig:gg} are nearly parallel, consistent with the amplitude from these higher-dimensional operators scaling as $\hat{s}^2$. 
The behavior of $C_{g,3}$ is more intricate: the signal flips sign around 1.8 TeV. Below this threshold, the negative interference from the $+$$-$$+$$-$ and $+$$-$$-$$+$ configurations dominates, whereas above this energy, the positive contribution from the $+$$+$$+$$+$ configuration overtakes.
For the normalized $d \sigma/(\sigma d\Delta y)$ distribution, the results for $C_u$ and $C_d$ are nearly indistinguishable across most of the $\Delta y$ range, with noticeable differences only in the low-$\Delta y$ regime. 
Compared to the Standard Model, the signals induced by $C_u$ and $C_d$ are more concentrated at low $\Delta y$, reflecting the enhancement in highly boosted configurations. The signals from the gluonic operators exhibit a similar pattern but distinct distributions, allowing them to be clearly separated.
\begin{figure}[h]
    \centering
    \includegraphics[width=0.75\linewidth]{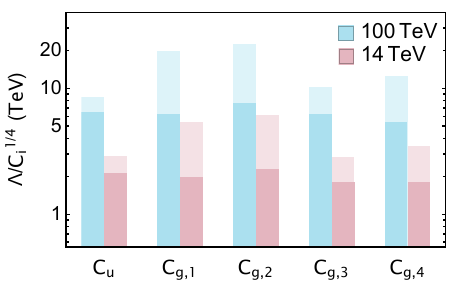}
    \caption{The 95\% C.L. reach for the effective scale $\Lambda/\left(C_i \right)^{1/4}$ of different operators. Light‑colored bars: individual operator bounds (one operator turned on at a time). Darker bars: marginalized projection with five operators varied simultaneously. Results assume an integrated luminosity of $L=3\,\mathrm{ab}^{-1}$ at 14 TeV and 100 TeV.
    }
    \label{fig:lambda}
\end{figure}

To set bounds on the Wilson coefficients (equivalently on the effective scale $\Lambda/( C_i^{(8)} )^{1/4}$), we perform a binned chi‑square analysis on the 2D differential distribution $\dd\sigma/(\dd \sqrt{\hat{s}} \, \dd\Delta y)$.
For the 14 TeV projections, the diphoton invariant mass $m_{\gamma\gamma}$ is binned from 200 GeV to 2 TeV at the step of 100 GeV, while $\Delta y$ is binned from 0 to 5 in intervals of 0.5. For the 100 TeV projections, we use $m_{\gamma\gamma}\in[1,6]\,$TeV with 500 GeV bin width and the same $\Delta y$ binning. Bins containing fewer than 10 expected SM events are discarded. We compute the chi‑square function for the $i$-th bin, 
\begin{equation}
    \Delta \chi_i^2 = \left( \dfrac{S_i}{B_i}\right)^2 \left(\dfrac{1}{B_i} +\delta_i^2 \right)^{-1} \ ,
\end{equation}
where $S_i$ and $B_i$ are the number of events for signal and backgrounds, and $\delta_i$ refers to the theoretical uncertainty from PDF scale variation. The total $\Delta \chi^2$ is obtained by summing over all bins.

\begin{figure}[!t]
    \centering
        \includegraphics[width=0.9\linewidth]{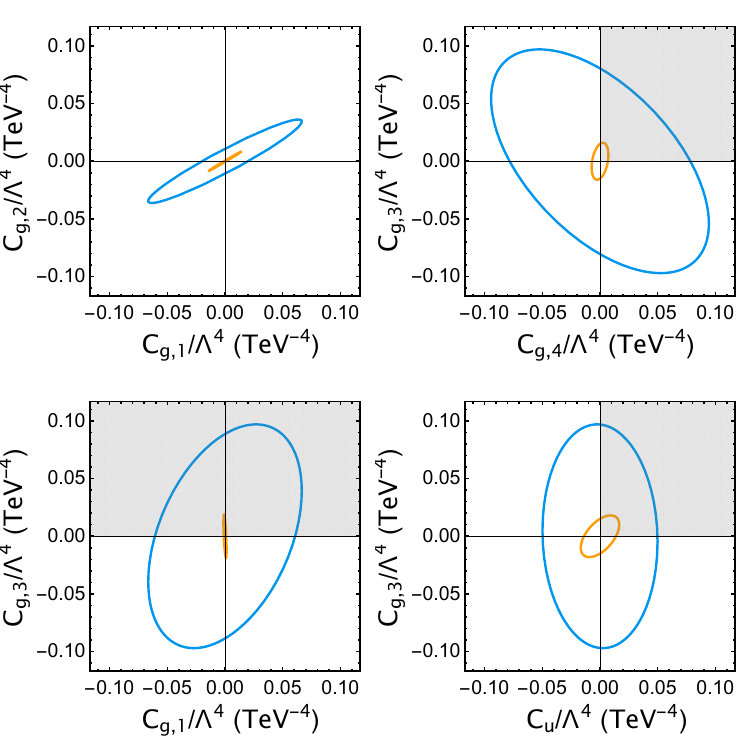}
    \caption{Projected 95\% C.L. contours in representative 2D planes of Wilson coefficients at the 14 TeV HL‑LHC. The blue contour is obtained when five coefficients (all except $C_d$) are floated and the orange contour shows the two‑parameter fit (only the two indicated coefficients are nonzero).}
    \label{fig:chisq}
\end{figure}

We require the new physics contributions to lie within the 95\% confidence interval of the SM prediction. The corresponding results are shown in Fig.~\ref{fig:lambda}. Since the contributions of $C_u$ and $C_d$ are nearly degenerate, we only turn on $C_u$ while marginalizing over the other four operators $C_i$.
The dark bar in the figure represents the result from the marginalized projection when turning on all operators, while the light bar corresponds to the sensitivity of individual operator turned on. 
For the marginalized projection, at 14 TeV HL-LHC, the effective probed scales are all around 2 TeV. At a future 100 TeV proton-proton collider with the same luminosity, the reach can extend to 5--8 TeV. When turning on individual operators, the projected sensitivity ranges from 3 to 5 TeV for 14 TeV HL-LHC, and from 7 to 20 TeV for the 100 TeV proton collider.

Furthermore, we show the correlation between different coefficient pairs on the 2D plane in Fig.~\ref{fig:chisq}.  
The blue curve is for the case of turning on five coefficients and the orange curve is for turning on two coefficients. The light shading indicates the region allowed by the positivity bounds for the corresponding coefficients. 
This indicates that the diphoton channel not only allows positivity bounds to be tested at high energy scales, but also enables SMEFT fittings to be combined with positivity constraints to obtain stronger bounds on the Wilson coefficients.

\section{Summary and Outlook}

In this paper, we examine the implications of positivity bounds on dimension-8 operators contributing to the diphoton production process at the LHC and future hadron colliders. We systematically identify all interactions involving two photons and either a quark or gluon pair, and show that the leading contributions start from dimension-8 operators.
The positivity bounds require the Wilson coefficients associated with quark-photon operators to be positive, while certain gluon-photon operators remain unconstrained due to the vanishing forward scattering amplitude. The quark-initiated operators modify the  $q\bar{q}\rightarrow \gamma\gamma$ process, whereas the gluonic operators contribute to the $g g \rightarrow \gamma\gamma$ channel. 
We analyze the resulting differential distributions and assess sensitivity using a binned chi‑square on ($m_{\gamma\gamma},\,\Delta y$). 
Two benchmark scenarios are considered: the 14 TeV HL-LHC and a prospective 100 TeV proton‑proton collider, both with an integrated luminosity of $L=3\,\mathrm{ab}^{-1}$. Working at interference order, we find that the 95\% C.L. exclusion limits for the effective scale typically reach $\sim2\,$TeV at the HL‑LHC and exceed $\sim5\,$TeV at a 100 TeV collider.
This provides a powerful probe of the positivity bounds in the quark and gluon sector, and thus tests the fundamental principles of the quantum field theory.

\section*{Acknowledgement}
We thank Jiayin Gu for helpful discussions.
 Z.L and K.-F.L are supported in part by U.S. Department of Energy Grant No. DESC0011842. Z.L. is supported in part by a Sloan Research Fellowship from the Alfred P. Sloan Foundation at the University of Minnesota.

\appendix

\section*{Appendix}

The projected sensitivities of the dimension-6 operators at HL-LHC. 

\begin{table}[th]
    \centering
    \begin{tabular}{|c|c|c|}
    \hline
        \rule{0pt}{2.4ex} Operator & Constraint & Signal \\
    \hline
        \rule{0pt}{2.4ex} $(\bar q\sigma^{\mu\nu}u)\tilde{\varphi} B_{\mu\nu}$ & $\Lambda>4.6\,\TeV$ \cite{daSilvaAlmeida:2019cbr} &  \\
    \cline{1-2}
        \rule{0pt}{2.4ex} $(\bar q\sigma^{\mu\nu}u)\tau^I\tilde{\varphi} W^I_{\mu\nu}$ & $\Lambda>4.2\,\TeV$ \cite{daSilvaAlmeida:2019cbr} & LO  \\
    \cline{1-2}
        \rule{0pt}{2.4ex} $(\bar q\sigma^{\mu\nu}d)\tilde{\varphi} B_{\mu\nu}$ & $\Lambda>4.2\,\TeV$ \cite{daSilvaAlmeida:2019cbr} &  $|M_{\rm d6}|^2\!\sim\!\frac{v^2s}{\Lambda^4}$ \\
    \cline{1-2}
        \rule{0pt}{2.4ex} $(\bar q\sigma^{\mu\nu}d)\tau^I\tilde{\varphi} W^I_{\mu\nu}$ & $\Lambda>3.9\,\TeV$ \cite{daSilvaAlmeida:2019cbr} &   \\
    \hline
        \rule{0pt}{2.4ex} $(\bar q\sigma^{\mu\nu}T^A u)\tilde{\varphi} G^A_{\mu\nu}$ & $\Lambda>2.6\,\TeV$ \cite{Hayreter:2013kba} & \multirow{2}{*}{NLO} \\
    \cline{1-2}
        \rule{0pt}{2.4ex}$(\bar q\sigma^{\mu\nu}T^A d)\tilde{\varphi} G^A_{\mu\nu}$ & $\Lambda>2.2\,\TeV$ \cite{Hayreter:2013kba} &  \\
    \hline
        \rule{0pt}{2.4ex}$\epsilon^{IJK}W^I W^J W^K$ & $\Lambda>10\,\TeV$ \cite{Bian:2015zha} & NLO \\
    \hline
        
    \end{tabular}
    \caption{Bounds on dimension-6 operators relevant for the diphoton production. All the bounds are recast and projected to HL-LHC assuming the dimensionless couplings $C=1$.  } 
    \label{tab:dim6}
\end{table}

\bibliography{references}

\end{document}